\documentclass[english]{paper}
\usepackage[T1]{fontenc}
\usepackage[latin9]{inputenc}
\usepackage{amsmath}
\usepackage{amssymb}

\makeatletter

\providecommand{\tabularnewline}{\\}

\newenvironment{lyxlist}[1]
{\begin{list}{}
{\settowidth{\labelwidth}{#1}
 \setlength{\leftmargin}{\labelwidth}
 \addtolength{\leftmargin}{\labelsep}
 }}
{\end{list}}

\usepackage{syntax}
\usepackage {url}
\usepackage [numbers]{natbib}

\makeatother

\usepackage{babel}
\begin{document}

\title{An Outline of Separation Logic}

\author{Abhishek Kr Singh and Raja Natrajan \\Tata Institute of Fundamental Research, Mumbai }
\maketitle
\begin{abstract}
Separation Logic is an effective Program Logic for proving programs
that involve pointers. Reasoning with pointers becomes difficult especially
when there is aliasing arising due to several pointers to a given
cell location. In this paper, we try to explore the problems with
aliasing through some simple examples and introduce the notion of
separating conjunction as a tool to deal with it. We introduce Separation
Logic as an extension of the standard Hoare Logic with the help pf
a programming language that has four pointer manipulating commands.
These commands perform the usual heap operations such as \emph{lookup,
update, allocation} and \emph{deallocation}. The new set of assertions
and axioms of Separation Logic is presented in a semi-formal style.
Examples are given to illustrate the unique features of the new assertions
and axioms. Finally the paper concludes with the proofs of some real
programs using the axioms of Separation Logic. 
\end{abstract}

\section{Introduction}

The goal of any Program Logic is to help in developing logically correct
programs without the need for debugging. \emph{Separation Logic} \cite{key-7}
is one such Program Logic. I can be seen as an extension to the standard
\emph{Hoare Logic} \cite{key-1}. The goal of this extension is to
simplify reasoning with programs that involve low level memory access,
such as \emph{pointers}.  
\begin{itemize}
\item Reasoning with pointers becomes very difficult because of the way
it interfere with the modular style of program development. 
\end{itemize}
Structured programming approaches provide the freedom to develop a
large program by splitting it into small modules. Hence, at any given
time a programmer can only concentrate on developing a particular
module against its specification. In the absence of pointer commands,
the proof of correctness of these small modules can easily be extended
to a global proof for the whole program.
\begin{itemize}
\item For example, consider the specification 
\end{itemize}
\begin{center}
$\{x=4\}$~~\texttt{x:= 8}~~$\{x=8\}$
\par\end{center}

in \emph{Hoare triple}. It claims that if execution of the command
\texttt{x:= 8} begins in a state where $x=4$ is true, then it ends
in a state where $x=8$ is true. This seems trivially true. In a similar
way, one can easily verify the validity of following specification:

\begin{center}
$\{x=4\wedge y=4\}$~~\texttt{x:= 8}~~ $\{x=8\wedge y=4\}$
\par\end{center}

Here, the proposition $y=4$ should remain true in the postcondition,
since the value of variable $y$ is not affected by the execution
of command \texttt{x:= 8}. 
\begin{itemize}
\item The reasoning above is an instance of a more general rule of Hoare
logic, called the \emph{rule of constancy}
\end{itemize}
\begin{center}
$\dfrac{\{p\}\,\,c\,\,\{q\}}{\{p\wedge r\}\,\,c\,\,\{q\wedge r\}}$
\par\end{center}

where, no free variable of $r$ is modified by $c.$ It is easy to
see how this rule follows from the usual meaning of the assignment
operator. Following sequence of state transition can be used to illustrate
the above idea:

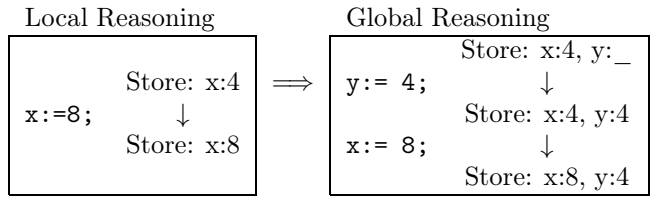
\begin{center}
\begin{tabular}{|cc|cc|cc|}
\multicolumn{2}{l}{Local Reasoning} & \multicolumn{2}{c}{} & \multicolumn{2}{l}{Global Reasoning}\tabularnewline
\cline{1-2} \cline{5-6} 
 &  &  &  &  & Store: x:4, y:\_\tabularnewline
 & Store: x:4 & \multicolumn{2}{c|}{$\Longrightarrow$} & \texttt{y:= 4;} & $\downarrow$\tabularnewline
\texttt{x:=8;} & $\downarrow$ &  &  &  & Store: x:4, y:4\tabularnewline
 & Store: x:8 &  &  & \texttt{x:= 8;} & $\downarrow$\tabularnewline
 &  &  &  &  & Store: x:8, y:4\tabularnewline
\cline{1-2} \cline{5-6} 
\end{tabular}
\par\end{center}

However, if the program modules uses data structures such as arrays,
linked lists or trees, which involves the addressable memory, then
extending the local reasoning is not so easy using the rule of constancy. 
\begin{itemize}
\item For example, consider a similar specification which involves mutation
of an array element
\end{itemize}
\begin{center}
$\{a\left[i\right]=4\wedge a\left[j\right]=4\}$ \texttt{a{[}i{]}:=
8 }$\{a\left[i\right]=8\wedge a\left[j\right]=4\}$
\par\end{center}

It is not a valid specification. Consider the case when $i=j$. Therefore
an extra clause $i\neq j$ is needed in the precondition to make it
a valid specification. To complicate the situation even further, consider
the following specification

\begin{center}
$\{a\left[i\right]=4\wedge a\left[j\right]=4\wedge a\left[k\right]=4\}$
\texttt{a{[}i{]}:= 8 }$\{a\left[i\right]=8\wedge a\left[j\right]=4\wedge a\left[k\right]=4\}$
\par\end{center}

In this case two more clauses i,e. $i\neq j$ and $i\neq k$ are needed
in the precondition to make it a valid specification. These are the
extra clauses, that a programmer often forgets to mention. However
these clauses are necessary since it assures that the three propositions
$a[i]=4$ , $a[j]=4$ and $a[k]=4$ refer to the mutually disjoint
portion of the heap memory and hence mutating one will not affect
the others. Thus, although $\{a\left[i\right]=4\}$\texttt{ a{[}i{]}:=
8} $\{a\left[i\right]=8\}$ is a valid specification, applying rule
of constancy in these cases can lead us to an invalid conclusion.

These kind of non-sharing is often assumed by programmers. However,
in classical logic non sharing need explicit mention, which results
in a program specification that looks clumsy.
\begin{itemize}
\item Separation logic deals with this difficulty by introducing a separating
conjunction, $P*Q$, which asserts that $P$ and $Q$ holds for disjoint
portions of the addressable memory. 
\item In this sense it is more close to the programmers way of reasoning. 
\end{itemize}
Since non sharing is default in separating conjunction, the above
specification can be written succinctly as

\begin{center}
$\{a\left[i\right]\mapsto4*a\left[j\right]\mapsto4*a\left[k\right]=4\}$
\texttt{a{[}i{]}:= 8 }$\{a\left[i\right]\mapsto8*a\left[j\right]\mapsto4*a\left[k\right]\mapsto4\}$
\par\end{center}

where, $p\mapsto e$ represents a single cell heap-let with $p$ as
domain and $e$ is the value stored at the address $p$. Thus the
assertion $a\left[i\right]\mapsto4*a\left[j\right]\mapsto4$ means
that $a\left[i\right]\mapsto4$ and $a\left[j\right]\mapsto4$ holds
on disjoint parts of the heap and hence $i\neq j$. Although, the
normal rule of constancy is no more valid, we have the following equivalent
rule called \emph{``frame rule''}

\begin{center}
$\dfrac{\{p\}\,\,c\,\,\{q\}}{\{p*r\}\,\,c\,\,\{q*r\}}$
\par\end{center}

where, no variable occurring free in $r$ is modified by $c$.

In this section, we have seen some specifications and their meanings
in a semi-formal way. We need to define these notions formally, before
we can discuss new assertions and other features of separation logic
more rigorously. Section \ref{sec:Background}, prepares this background
by defining a language $\mathcal{L}$ and introducing Hoare Logic.
Section \ref{sec:New-Assertions-and}, introduces the new forms of
assertions in Separation Logic. It also extends axioms of Hoare Logic
to include some new axioms for reasoning with pointers. In Section
\ref{sec:Annotated-proofs}, we describe the idea of annotated proofs
and present the proof of an in-place list reversal program using the
axioms of Separation logic. 

\section{\label{sec:Background}Background}

In this section, we fix a language $\mathcal{L}$ by defining its
syntax and semantics. A subset $S$ of the language $\mathcal{L}$
is then used to introduce the axioms of Hoare Logic for commands that
doesn't involve pointers. We now describe the structure and meaning
of various commands in the language $\mathcal{L}$:
\begin{itemize}
\item \textbf{Skip.}\\
Command: $\texttt{skip}$ \\
Meaning: The execution has no effect on the state of computation.
\item \textbf{Assignment.} \\
Command: $\texttt{x:= e}$\\
Meaning: The command changes the state by assigning the value of term
$e$ to the variable $x$. 
\item \textbf{Sequencing.}\\
Command: $C_{1};\dots;C_{n}$\\
Meaning: The commands $C_{1},\dots,C_{n}$ are executed in that order.
\item \textbf{Conditional}.\\
Command: $\texttt{if b then \ensuremath{C_{1}}else \ensuremath{C_{2}}}$\\
Meaning: If the boolean expression $b$ evaluates to \textbf{true}
in the present state, then $C_{1}$ is executed. If $b$ evaluates
to \textbf{false}, then $C_{2}$ is executed.
\item \textbf{While-Loop}. \\
Command: $\texttt{while b do \ensuremath{C}}$\\
Meaning: If the boolean expression $b$ evaluates to \textbf{false}
then nothing is done. If $b$ evaluates to \textbf{true} in the present
state, then $C$ is executed and the $\texttt{while }$command is
then repeated. Hence, $C$ is repeatedly executed until the value
of $b$ becomes \textbf{false}. 
\item \textbf{Allocation}.\\
Command: $\texttt{x:=cons(\ensuremath{e_{1},\dots,e_{n}})}$\\
Meaning: The command $\texttt{x:=cons(\ensuremath{e_{1},\dots,e_{n}})}$
reserves \emph{$n$} consecutive cells in the memory initialized to
the values of $\ensuremath{e_{1},\dots,e_{n}}$, and saves in $x$
the address of first the cell. Note that, for the successful execution
of this command the addressable memory must have $n$ uninitialized
and consecutive cells available. 
\item \textbf{Lookup.}\\
Command: $\texttt{ x:= [e] }$\\
Meaning: It saves the value stored at location $e$ in the variable
$x$. Again for the successful execution of this command location
$e$ must have been initialized by some previous command of the program.
Otherwise, the execution will abort.
\item \textbf{Mutation.}\\
Command:$\texttt{[e]:= e'}$\\
Meaning: The command $\texttt{[e]:= e'}$, stores the value of expression
$e'$ at the location $e$. Again, for this to happen, location $e$
must be an active cell of the addressable memory.
\item \textbf{Deallocation.}\\
Command: $\texttt{free (e)}$ \\
Meaning: The instruction $\texttt{free (e)}$ , deallocates the cell
at the address $e$. If $e$ is not an active cell location, then
the execution of this command shall abort. 
\end{itemize}

\subsection{Formal Syntax}

The structure of commands in the language $\mathcal{L}$ can also
be described by the following abstract syntax:

\paragraph{Syntax of Command} 
\begin{grammar}
<cmd> ::= `skip' 
\alt <var> := <aexp>   
\alt <cmd> `;' <cmd>
\alt `if' <bexp> `then' <cmd> `else' <cmd>
\alt `while' <bexp> `do' <cmd>
\alt <var> := `cons' (<aexp>,..., <aexp>)
\alt <var> := [<aexp>]
\alt [<aexp>] := <aexp>
\alt `free' (<aexp>)
\end{grammar} Where \emph{aexp} and \emph{bexp} stands for arithmetic and boolean expressions respectively. The syntax of these are as follows: 
\begin{grammar}
<aexp>	::= \textbf{int}	
\alt <var> 	
\alt <aexp> + <aexp>	
\alt <aexp> - <aexp>	
\alt <aexp>  $\times$  <aexp>
\end{grammar}

\begin {grammar}
<bexp>	::= \textbf{true} | \textbf{false} 
\alt <aexp> = <aexp>
\alt $\neg$ <bexp>
\alt <bexp> $\wedge$ <bexp>
\alt <bexp> $\vee$ <bexp>
\alt <bexp> $\Rightarrow$ <bexp>
\end{grammar}

\subsection{Formal Semantics}

The formal semantics of a programming language can be specified by
assigning meanings to its individual commands. A natural way of assigning
meaning to a command is by describing the effect of its execution
on the \emph{state} of computation. The State of a computation can
be described by its two components, \emph{store} and \emph{heap}. 
\begin{itemize}
\item Store, which is sometimes called stack, contains the values of local
variables. Heap maintains the information about the contents of active
cell locations in the memory. More precisely, both of them can be
viewed as partial functions of the following form:
\end{itemize}
\begin{center}
\begin{tabular}{lcccr}
\textsf{Heaps $\triangleq$Location $\rightharpoonup$ Int} &  &  &  & \textsf{Stores $\triangleq$Variables $\rightharpoonup$Int}\tabularnewline
\end{tabular}
\par\end{center}
\begin{itemize}
\item Note that the notations, \textsf{\textbf{cons}} and $[-]$, which
refer to the heap memory, are absent in the syntax of \emph{aexp}.
Therefore, the evaluation of an arithmetic or boolean expression depends
only on the contents of the store at any given time. We use the notation
$s\models e\Downarrow v$ to assert that the expression $e$ evaluates
to $v$ with respect to the content of store $s$. For example, let
$s=\{(x,2),(y,4),(z,6)\}$ then $s\models x\times(y+z)\Downarrow20$
. For our discussion, we assume that this evaluation relation is already
defined. 
\end{itemize}

\subsubsection*{Operational Semantics:}

We now define a transition relation, represented as $\left\langle c,(s,h)\right\rangle \,\,\rightarrowtail(s',h')$,
between states. It asserts that, if the execution of command $c$
starts in a state $(s,h)$, then it will end in the state $(s',h')$.
The following set of rules, SEMANTICS-I and -II, describes the operational
behavior of every command in the language $\mathcal{L}$, using the
transition relation $\rightarrowtail$. 

\begin{center}
\begin{tabular}{|cl|}
\multicolumn{2}{c}{SEMANTICS- I (Commands without pointers)}\tabularnewline
\hline 
 & \tabularnewline
Skip & \texttt{$\left\langle skip,(s,h)\right\rangle $}~~$\rightarrowtail(s,h)$\tabularnewline
 & \tabularnewline
Asign & $\dfrac{s\models e\Downarrow v}{\left\langle \text{\texttt{x:= e }},\,\,(s,h)\right\rangle \,\,\,\rightarrowtail(s[x:v],h)}$\tabularnewline
 & \tabularnewline
Seq & $\dfrac{\left\langle c_{1},\,\,(s,h)\right\rangle \,\,\,\rightarrowtail(s',h')}{\left\langle c_{1};c_{2},\,\,(s,h)\right\rangle \,\,\,\rightarrowtail\,\,\left\langle c_{2},\,\,(s',h')\right\rangle }$\tabularnewline
 & \tabularnewline
If-T & $\dfrac{s\models e\Downarrow\text{\textbf{true}}\,\,\,\,\,\,\,\,\,\,\,\left\langle c_{1},\,\,(s,h)\right\rangle \,\,\rightarrowtail\,\,(s',h')}{\left\langle \text{\texttt{if} }\,e\,\text{ \texttt{then} }\,c_{1}\,\text{\texttt{ else} }c_{2},\,\,(s,h)\right\rangle \,\,\,\rightarrowtail\,\,(s',h')}$\tabularnewline
 & \tabularnewline
If-F & $\dfrac{s\models e\Downarrow\text{\textbf{false}}\,\,\,\,\,\,\,\,\,\,\,\left\langle c_{2},\,\,(s,h)\right\rangle \,\,\rightarrowtail\,\,(s',h')}{\left\langle \text{\texttt{if} }\,e\,\text{ \texttt{then} }\,c_{1}\,\text{\texttt{ else} }c_{2},\,\,(s,h)\right\rangle \,\,\,\rightarrowtail\,\,(s',h')}$\tabularnewline
 & \tabularnewline
W-F & $\dfrac{s\models e\Downarrow\textbf{false}}{\left\langle \texttt{while }e\texttt{ do }c,\,\,(s,h)\right\rangle \,\,\rightarrowtail\,\,(s,\,h)}$\tabularnewline
 & \tabularnewline
\multicolumn{2}{|r|}{W-T$\dfrac{s\models e\Downarrow\textbf{true}\,\,\,\,\left\langle c,\,\,(s,h)\right\rangle \,\rightarrowtail(s',h')\,\,\,\,\left\langle \texttt{while }e\texttt{ do }c,(s',h')\right\rangle \,\,\rightarrowtail\,\,(s'',h'')}{\left\langle \texttt{while }e\texttt{ do }c,\,\,(s,h)\right\rangle \,\,\rightarrowtail\,\,(s'',h'')}$}\tabularnewline
 & \tabularnewline
\hline 
\multicolumn{1}{c}{} & \multicolumn{1}{l}{}\tabularnewline
\end{tabular}
\par\end{center}

\begin{center}
\begin{tabular}{|rl|}
\multicolumn{2}{c}{SEMANTICS-II (Commands with pointers)}\tabularnewline
\hline 
 & \tabularnewline
Alloc & $\dfrac{s\models e_{1}\Downarrow v_{1},\dots,s\models e_{n}\Downarrow v_{n}\,\,\,\,\,\,\,\,\,l,\dots,l+n-1\in\texttt{Locations}-\texttt{dom }h}{\left\langle \texttt{x:=cons(\ensuremath{e_{1},\dots,e_{n}}),\,\,\ensuremath{(s,h)}}\right\rangle \,\,\rightarrowtail\,\,(s[x:l],\,\,h[l:v_{1},\dots,l+n-1:v_{n}])}$\tabularnewline
 & \tabularnewline
Look & $\dfrac{s\models e\Downarrow v\,\,\,\,\,\,\,\,\,v\in\texttt{dom }h}{\left\langle \texttt{x:= [\ensuremath{e}]},\,\,(s,h)\right\rangle \,\,\rightarrowtail\,\,(s[x:h(v)],h)}$~~~~$\dfrac{s\models e\Downarrow v\,\,\,\,\,\,\,\,\,v\notin\texttt{dom }h}{\left\langle \texttt{x:= [\ensuremath{e}]},\,\,(s,h)\right\rangle \,\,\rightarrowtail\,\,\textbf{abort}}$\tabularnewline
 & \tabularnewline
Mut & $\dfrac{s\models e\Downarrow v\,\,\,\,\,\,\,\,\,v\in\texttt{dom }h\,\,\,\,\,\,\,\,s\models e'\Downarrow v'}{\left\langle \texttt{[\ensuremath{e}]:= \ensuremath{e'}},\,\,(s,h)\right\rangle \,\,\rightarrowtail\,\,(s,\,h[v:v'])}$~~~$\dfrac{s\models e\Downarrow v\,\,\,\,\,\,\,\,\,v\notin\texttt{dom }h}{\left\langle \texttt{[\ensuremath{e}]:= \ensuremath{e'}},\,\,(s,h)\right\rangle \,\,\rightarrowtail\,\,\textbf{abort}}$\tabularnewline
 & \tabularnewline
\multicolumn{2}{|r|}{Free$\dfrac{s\models e\Downarrow v\,\,\,\,\,\,\,\,\,v\in\texttt{dom }h}{\left\langle \texttt{free }(e),(s,h)\right\rangle \,\,\rightarrowtail\,\,(s,\,h\rceil(\texttt{dom }h\,-\{v\}))}$~~~$\dfrac{s\models e\Downarrow v\,\,\,\,\,\,\,\,\,v\notin\texttt{dom }h}{\left\langle \texttt{free }(e),(s,h)\right\rangle \,\,\rightarrowtail\,\,\textbf{abort}}$}\tabularnewline
 & \tabularnewline
\hline 
\end{tabular}
\par\end{center}

\begin{flushleft}
Where, $f[x:v]$ represents a function that maps $x$ to $v$ and
all other argument $y$ in the domain of $f$ to $f\,y$. Notation
$f\rceil A$ is used to represent the restriction of function $f$
to the domain $A$. 
\par\end{flushleft}
\begin{itemize}
\item An important feature of the Language $\mathcal{L}$ is that, any attempt
to refer to an unallocated address causes the program execution to
\textbf{abort}. For example, consider the following sequence of commands,
\end{itemize}
\begin{center}
\begin{tabular}{lclc}
 &  &  & Store~~ x:0, y:0\tabularnewline
 &  &  & Heap~~ empty\tabularnewline
Allocation &  & \texttt{x:= cons(1,2);} & $\downarrow$\tabularnewline
 &  &  & Store~~x:10, y:0\tabularnewline
 &  &  & Heap~~10:1, 11:2\tabularnewline
Lookup &  & \texttt{y:= {[}x{]};} & $\downarrow$\tabularnewline
 &  &  & Store~~x:10, y:1\tabularnewline
 &  &  & Heap~~10:1, 11:2\tabularnewline
Deallocation &  & \texttt{free(x+1);} & $\downarrow$\tabularnewline
 &  &  & Store~~x:10, y:1\tabularnewline
 &  &  & Heap~~10:1\tabularnewline
Mutation &  & \texttt{{[}x+1{]}:= y;} & $\downarrow$\tabularnewline
 &  &  & \textbf{abort}\tabularnewline
\end{tabular}
\par\end{center}

here, an attempt to mutate the content of address $11$ causes the
execution to \textbf{abort}, because this location was deallocated
by the previous instruction.

\subsection{Hoare Triple}

The operational semantics of language $\mathcal{L}$ can be used to
prove any valid specification of the form $\left\langle c,(s,h)\right\rangle \,\,\overset{*}{\rightarrowtail}\,(s',h')$.
However, this form of specification is not the most useful one. Usually,
we do not wish to specify programs for single states. Instead, we
would like to talk about a set of states and how the execution may
transform that set. This is possible using a Hoare triple $\{p\}c\{q\}$, 
\begin{itemize}
\item Informally, it says that if the execution of program $c$ begins in
a state that satisfies proposition $p$ then if the execution terminates
it will end in a state that satisfies $q$. 
\end{itemize}
where $p$ and $q$ are assertions that may evaluate to either \textbf{true}
or \textbf{false} in a given state. We will use notation $[\![p]\!]\,\,s\,h$,
to represent the value to which $p$ evaluates, in the state $(s,h)$.
Therefore, 
\begin{itemize}
\item $\{p\}c\{q\}$ holds iff $\forall(s,h)\in\textsf{States},\,[\![p]\!]\,s\,h$
$\implies$\\
 $\neg(\left\langle c,(s,h)\right\rangle \overset{*}{\rightarrowtail}\textbf{abort})\,\wedge\,(\forall(s',h')\in\textsf{States, (\ensuremath{\left\langle c,(s,h)\right\rangle \overset{*}{\rightarrowtail}(s',h')})\ensuremath{\implies}\ensuremath{[\![q]\!]}}s'h').$ 
\end{itemize}
Now, we can use Hoare triples, to give rules of reasoning for every
individual commands of the language $\mathcal{L}$. It is sometimes
also called the \emph{axiomatic semantics} of the language. These
rules in a way give an alternative semantics to the commands. 

\subsubsection*{Axiomatic Semantics: }

Consider the following set of axioms (AXIOM-I) and structural rules
for reasoning with commands, that does not use pointers. Here, $Q[e/x]$
represents the proposition $Q$ with every free occurrence of $x$
replaced by the expression $e$, $Mod(c)$ represents the set of variables
modified by $c$, and $Free(R)$ represents the set of free variables
in $R$.

\begin{center}
\begin{tabular}{|rl|}
\multicolumn{2}{c}{AXIOMS-I}\tabularnewline
\hline 
 & \tabularnewline
skip & $\forall P:\textsf{Assert, }\{P\}\texttt{skip}\{P\}$\tabularnewline
 & \tabularnewline
assign & $\dfrac{}{\{Q[e/x]\}\texttt{x:=e}\{Q\}}$\tabularnewline
 & \tabularnewline
seq & $\dfrac{\{P\}c_{1}\{Q\}\,\,\,\,\,\,\,\{Q\}c_{2}\{R\}}{\{P\}c_{1};c_{2}\{R\}}$\tabularnewline
 & \tabularnewline
if &  $\dfrac{\{P\wedge e\}c_{1}\{Q\}\,\,\,\,\,\,\,\,\{P\wedge\neg e\}c_{2}\{Q\}}{\{P\}\texttt{if }e\texttt{ then }c_{1}\texttt{ else }c_{2}\{Q\}}$\tabularnewline
 & \tabularnewline
while & $\dfrac{\{I\wedge e\}c\{I\}}{\{I\}\texttt{while }e\texttt{ do }c\{I\wedge\neg e\}}$\tabularnewline
 & \tabularnewline
\hline 
\multicolumn{1}{r}{} & \multicolumn{1}{l}{}\tabularnewline
\end{tabular}
\par\end{center}

\begin{center}
\begin{tabular}{|rl|}
\multicolumn{2}{c}{STRUCTURAL RULES}\tabularnewline
\hline 
 & \tabularnewline
conseq & $\dfrac{P\implies P'\,\,\,\,\{P'\}c\{Q'\}\,\,\,\,Q'\implies Q}{\{P\}c\{Q\}}$\tabularnewline
 & \tabularnewline
extractE & $\dfrac{\{P\}c\{Q\}}{\{\exists x.P\}c\{\exists x.Q\}}$$x\notin Free(c)$\tabularnewline
 & \tabularnewline
var-sub & $\dfrac{\{P\}c\{Q\}}{(\{P\}c\{Q\})[E_{1}/x_{1},\dots,E_{k}/x_{k}]}$%
\begin{tabular}{c}
$x_{i}\in Mod(c)\text{ implies }$\tabularnewline
$\forall j\neq i,E_{i}\notin Free(E_{j})$\tabularnewline
\end{tabular}\tabularnewline
 & \tabularnewline
constancy & $\dfrac{\{P\}c\{Q\}}{\{P\wedge R\}c\{Q\wedge R\}}Mod(c)\cap Free(R)=\phi$\tabularnewline
 & \tabularnewline
\hline 
\end{tabular}
\par\end{center}
\begin{itemize}
\item For the subset of language $\mathcal{L}$, that does not use pointers,
these axioms and structural rules are known to be sound as well as
complete \cite{key-10} with respect to the operational semantics. 
\item The benefit of using these axioms is that, we can work on a more abstract
level specifying and proving program correctness in an axiomatic way
without bothering about low level details of states. 
\item Note that, the last four commands of the language $\mathcal{L}$,
which manipulate pointers, are different from the normal variable
assignment command. Their right hand side is not an expression. Therefore,
Hoare-assign rule is not applicable on them.
\end{itemize}

\subsubsection*{Array Revisited:}

Now, let us go back to the same array assignment problem which we
discussed in the introduction. Let $Q$ be the postcondition for the
command \texttt{a{[}i{]}:= 8}, where \texttt{$\texttt{a[i]}$} refers
to the $i^{th}$ element of array. 
\begin{itemize}
\item In our language $\mathcal{L}$ this is same as command \texttt{{[}a+i{]}:=
8}, however, for convenience we will continue with the usual notation
of array. 
\end{itemize}
The command \texttt{a{[}i{]}:= 8} looks similar to the variable assignment
command. But, we can not apply Hoare-assign rule to get $Q[8/\texttt{a[i]}]$
as weakest precondition. One should not treat \texttt{a{[}i{]}} as
a local variable, because the assertion $Q$ may contain references
such as \texttt{a{[}j{]}} that may or may not refer to\texttt{ a{[}i{]}}.
Instead, we can model the above command as \texttt{a:=update(a,i,8)},
where \texttt{update(a,i,8)}$[i]=8$ and \texttt{update(a,i,8)}$[j]=$a$[j]$
for $j\neq i$. 
\begin{itemize}
\item That is, the effect of executing \texttt{a{[}i{]}:= v} is same as
assigning variable \texttt{a} an altogether new array value \texttt{``update
(a,i,v)''.} 
\item In this way, \texttt{a} is acting like a normal variable, hence we
have the following rule for array assignment,
\end{itemize}
\begin{center}
array-assign$\dfrac{}{\{Q[\texttt{update(a,i,v)/a}]\}\texttt{ a[i]:= v }\{Q\}}$
\par\end{center}

Let us try to prove the following specification using above rule,

$\{i\neq j\wedge a\left[i\right]=4\wedge a\left[j\right]=4\}$ \texttt{a{[}i{]}:=8
}$\{a\left[i\right]=8\wedge a\left[j\right]=4\wedge i\neq j\}$.

Let $P$ =$\{i\neq j\wedge a\left[i\right]=4\wedge a\left[j\right]=4\}$
and $Q$ = $\{a\left[i\right]=8\wedge a\left[j\right]=4\wedge i\neq j\}$.
\\
\begin{tabular}{rrl}
Then we have, & \multicolumn{2}{l}{$Q$ {[}\texttt{update(a,i,8)/a}{]}}\tabularnewline
 &  & =$\{\texttt{update(a,i,8)}\left[i\right]=8\wedge\texttt{update(a,i,8)}\left[j\right]=4\wedge i\neq j\}$\tabularnewline
 &  & =$\{8=8\wedge\,a[j]=4\wedge i\neq j\}$\tabularnewline
 &  & =$\{a[j]=4\,\wedge\,\,i\neq j\}$\tabularnewline
\end{tabular}

Thus,

\begin{center}
\begin{tabular}{c}
$\dfrac{P\implies Q[\texttt{update(a,i,8)/a}]\,\,\,\,\,\,\,\{Q[\texttt{update(a,i,8)/a}]\}\texttt{a[i]:=8 }\{Q\}}{\{P\}\texttt{a[i]:=8}\{Q\}}$\tabularnewline
\end{tabular}
\par\end{center}

Hence, we have a correct rule for deducing valid specifications about
array assignments. However, the approach looks very clumsy. 
\begin{itemize}
\item We still need to fill every minute detail of index disjointness in
the specification. 
\item Moreover, it seems very artificial to interpret a local update to
an array cell as a global update to the whole array. At least, it
is not the programmer's way of understanding an array element update. 
\item The idea of separation logic is to embed the principle of such local
actions in the separating conjunction. It helps in keeping the specifications
succinct by avoiding the explicit mention of memory disjointness. 
\end{itemize}

\section{\label{sec:New-Assertions-and}New Assertions and Inference Rules}

In this section, we present the axioms corresponding to the pointer
manipulating commands. The set of assertions, which we use for this
purpose, goes beyond the predicates used in the Hoare- Logic. Following
is the syntax of the new assertions,

\begin{grammar}
<assert> ::=   \dots 
\alt \textbf{emp}
\alt <aexp> $\mapsto$ <aexp>   
\alt <assert> $*$ <assert>
\alt <assert> $-*$ <assert>
\end{grammar}

It is important to note that the meaning of these new assertions depends
on both the store and the heap.

\begin{tabular}{cl}
- & $\textbf{emp}$\tabularnewline
 & The heap is empty.\tabularnewline
- & $e_{1}\mapsto e_{2}$\tabularnewline
 & The heap contains a single cell, at address $e_{1}$ with contents
$e_{2}$.\tabularnewline
- & $p_{1}*p_{2}$\tabularnewline
 & The heap can be split into two disjoint parts in which $p_{0}$and
$p_{1}$ \tabularnewline
 & hold respectively.\tabularnewline
- & $p_{1}-\!\!*\,\,p_{2}$\tabularnewline
 & If the current heap is extended with a disjoint part in which $p_{1}$ \tabularnewline
 & holds, then $p_{2}$ holds for the extended heap.\tabularnewline
 & \tabularnewline
\end{tabular}

For convenience, we introduce few more notations for the following
assertions: 

\begin{center}
\begin{tabular}{rcl}
$e\mapsto-$ & $\triangleq$ & $\exists x.\,\,e\mapsto x$ where $x$ is not free in $e$\tabularnewline
 &  & \tabularnewline
$e\hookrightarrow e'$ & $\triangleq$ & $e\mapsto e'\,*\,\textbf{true}$\tabularnewline
 &  & \tabularnewline
$e\mapsto e_{1},\dots,e_{n}$ & $\triangleq$ & $e\mapsto e_{1}\,*\dots*\,e+n-1\mapsto e_{n}$\tabularnewline
 &  & \tabularnewline
$e\hookrightarrow e_{1},\dots,e_{n}$ & $\triangleq$ & $e\hookrightarrow e_{1}\,*\dots*\,e+n-1\hookrightarrow e_{n}$\tabularnewline
 &  & i ff $e\mapsto e_{1},\dots,e_{n}\,\,*\,\,\textbf{true}$.\tabularnewline
\end{tabular}
\par\end{center}

We now consider a simple example to explore some of the interesting
features of separating conjunction. Let $h_{1}=\{(sx,1)\}$ and $h_{2}=\{(sy,2)\}$
be heaps where $s$ is a store such that $sx\neq sy$. Then, one can
verify the following

\begin{center}
\begin{tabular}{cccc}
1. & $[\![x\mapsto1*y\mapsto2]\!]\,\,s\,h$ & iff & $h=h_{1}.h_{2}$\tabularnewline
2. & $[\![x\mapsto1\wedge x\mapsto1]\!]\,\,s\,h$ & iff & $h=h_{1}$\tabularnewline
3. & $[\![x\mapsto1*x\mapsto1]\!]\,\,s\,h$ & iff & $\textbf{false}$\tabularnewline
4. & $[\![x\mapsto1\vee y\mapsto2]\!]\,\,s\,h$ & iff & $h=h_{1}\text{or }h=h_{2}$\tabularnewline
5. & $[\![x\mapsto1*(x\mapsto1\vee y\mapsto2)]\!]\,\,s\,h$ & iff & $h=h_{1}.h_{2}$\tabularnewline
6. & $[\![(x\mapsto1\vee y\mapsto2)*(x\mapsto1\vee y\mapsto2)]\!]\,\,s\,h$ & iff & $h=h_{1}.h_{2}$\tabularnewline
7. & $[\![(x\mapsto1*y\mapsto2*(x\mapsto1\vee y\mapsto2)]\!]\,\,s\,h$ & iff & $\textbf{false}$\tabularnewline
\end{tabular}
\par\end{center}

Assertions 2 and 3 illustrates the difference between the behavior
of the classical conjunction and the separating conjunction. Both
the occurrence of $x\mapsto1$ in the assertion $[\![x\mapsto1*x\mapsto1]\!]\,\,s\,h$
is true for the same singleton heap $h_{1}$. Hence any heap $h$,
can never be split into two disjoint parts that satisfies $x\mapsto1$.
One can also compare assertions 3 and 6, which looks similar in structure
but have different behaviors. 

The separating conjunction obeys commutative, associative, and some
distributive as well as semi-distributive laws. The assertion \textbf{emp}
behaves like a neutral elements. Most of these properties are contained
in the following axiom schemata. Note the use of one directional implications
in $(p_{1}\wedge p_{2})*q\implies(p_{1}*q)\wedge(p_{2}*q)$ and $(\forall x.p)*q\implies\forall x.(p*q)$.

\begin{center}
\begin{tabular}{rcl}
$p*\textbf{emp}$ & $\iff$ & $p$\tabularnewline
$p_{1}*p_{2}$ & $\iff$ & $p_{2}*p_{1}$\tabularnewline
$(p_{1}*p_{2})*p_{3}$ & $\iff$ & $p_{1}*(p_{2}*p_{3})$\tabularnewline
$(p_{1}\vee p_{2})*q$ & $\iff$ & $(p_{1}*q)\vee(p_{2}*q)$\tabularnewline
$(p_{1}\wedge p_{2})*q$ & $\implies$ & $(p_{1}*q)\wedge(p_{2}*q)$\tabularnewline
$(\exists x.p)*q$ & $\iff$ & $\exists x.(p*q)$ where $x$ is not free in $q$\tabularnewline
$(\forall x.p)*q$ & $\implies$ & $\forall x.(p*q)$ where $x$ is not free in q\tabularnewline
\end{tabular}
\par\end{center}

\subsubsection*{New Axioms for pointers:}

The axioms needed to reason about pointers are given below. There
is one axiom for every individual command. 

\begin{center}
\begin{tabular}{|rl|}
\multicolumn{2}{c}{AXIOMS-II}\tabularnewline
\hline 
 & \tabularnewline
alloc & $\dfrac{}{\{x=X\wedge\textbf{emp}\}\texttt{x:= cons(\ensuremath{e_{1},\dots,e_{k}})}\{x\mapsto\ensuremath{e_{1}[X/x],\dots,e_{k}[X/x]}\}}$\tabularnewline
 & \tabularnewline
lookup & $\dfrac{}{\{e\mapsto v\wedge x=X\}\texttt{ x:=[e] }\{x=v\wedge e[X/x]\mapsto v\}}$\tabularnewline
 & \tabularnewline
mut & $\dfrac{}{\{e\mapsto-\}\texttt{ [e]:= e\ensuremath{'}}\{e\mapsto e'\}}$\tabularnewline
 & \tabularnewline
free & $\dfrac{}{\{e\mapsto-\}\texttt{free }(e)\{\textbf{emp}\}}$\tabularnewline
 & \tabularnewline
\hline 
\end{tabular}
\par\end{center}
\begin{lyxlist}{00.00.0000}
\item [{allocate}] The first axiom, called \emph{alloc}, uses variable
$X$ in its precondition to record the value of $x$ before the command
is executed. It says that if execution begins with empty heap and
a store with $x=X$ then it ends with $k$ contiguous heap cells having
appropriate values.
\item [{lookup}] The second axiom, called \emph{lookup}, again uses $X$
to refer to the value of $x$ before execution. It asserts that the
content of heap is unaltered. The only change is in the store where
the new value of $x$ is modified to the value at old location $e$.
\item [{mutate}] The third axiom, called \emph{mut}, says that if $e$
points to something beforehand, then it points to $e'$ afterward.
This resembles the natural semantics of Mutation.
\item [{free}] The last axiom, called \emph{free}, says that if $e$ is
the only allocated memory cell before execution of the command then
in the resulting state there will be no active cell. Note that, the
singleton heap assertion is necessary in precondition to assure \textbf{emp}
in the postcondition.
\item [{frame}] The last rule among the structural rules, called \emph{frame},
says that one can extend local specifications to include any arbitrary
claims about variables and heap segments which are not modified or
mutated by $c$. The frame rule can be thought as an replacement to
the rule of constancy when pointers are involved.
\end{lyxlist}
\begin{center}
\begin{tabular}{|rl|}
\multicolumn{2}{c}{STRUCTURAL RULES-II}\tabularnewline
\hline 
 & \tabularnewline
conseq & $\dfrac{P\implies P'\,\,\,\,\{P'\}c\{Q'\}\,\,\,\,Q'\implies Q}{\{P\}c\{Q\}}$\tabularnewline
 & \tabularnewline
extractE & $\dfrac{\{P\}c\{Q\}}{\{\exists x.P\}c\{\exists x.Q\}}$$x\notin Free(c)$\tabularnewline
 & \tabularnewline
var-sub & $\dfrac{\{P\}c\{Q\}}{(\{P\}c\{Q\})[E_{1}/x_{1},\dots,E_{k}/x_{k}]}$%
\begin{tabular}{c}
$x_{i}\in Mod(c)\text{ implies }$\tabularnewline
$\forall j\neq i,E_{i}\notin Free(E_{j})$\tabularnewline
\end{tabular}\tabularnewline
frame & $\dfrac{\{p\}c\{q\}}{\{p*r\}c\{q*r\}}Mod(c)\cap Free(r)=\phi$\tabularnewline
 & \tabularnewline
\hline 
\end{tabular}
\par\end{center}
\begin{itemize}
\item Note that the expressions are intentionally kept free from the \texttt{cons}
and \texttt{{[}-{]}} operators. The reason for this restriction is
that, the power of the above proof system strongly depends upon the
ability to use expression in place of variables in an assertion. 
\item In particular, a tautology should remain a valid assertion on replacing
variables with expressions. 
\item If we could substitute \texttt{cons} $(e_{1},e_{2})$ for $x$ in
the tautology $x=x$, we obtain \texttt{cons$(e_{1},e_{2})=$cons}$(e_{1},e_{2})$,
which may not be a valid assertion if we wish to distinguish between
different addresses having the same content. 
\item Similarly \texttt{{[}-{]}} can not be used in expressions because
of the way it interact with separating conjunction. For example consider
substituting $[e]$ for $x$ and $y$ in the tautology $x=x\,*\,y=y$.
Clearly, $[e]=[e]\,\,*\,\,[e]=[e]$ is not a valid assertion.
\item Note that each axiom mentions only the portion of heap accessed by
the corresponding command. In this sense the axioms are local. Hence
a separate rule, called frame rule, is needed to extend these local
reasoning for a global context.
\item These axioms can easily be proved to be sound with respect to the
operational semantics of the language $\mathcal{L}$. 
\item Moreover, Yang in his thesis \cite{key-6} has shown that all valid
Hoare triples can be derived using the above collection of axioms
and the structural rules. In this sense these set of axioms and structural
rules are also complete.
\end{itemize}

\subsubsection*{Derived Rules:}

Although the small set of rules discussed so far is complete, it is
not practical. Proving a specification using this small set of axioms
often requires extensive invocation of the structural rules. Therefore,
it is good to have some derived rules that can be applied at once
in common situations. We now list some other useful rules that can
be derived from the natural semantics of the language $\mathcal{L}$.
A more detailed discussion about these rules can be found in \cite{key-9}.
Note that $x,\,\,x'$ and $X$ are all distinct variables.
\begin{itemize}
\item Assignment

\begin{itemize}
\item Forward reasoning \\
$\dfrac{}{\{x=X\}\texttt{ x:= e }\{x=e[X/x]\}}$
\item Floyd's forward running axiom\\
 $\dfrac{}{\{P\}\texttt{ x:= e }\{\exists x'.x=e[x'/x]\wedge P[x'/x]\}}$
\end{itemize}
\item Mutation

\begin{itemize}
\item Global reasoning\\
 $\dfrac{}{\{(e\mapsto-)*r\}\texttt{ [e]:= e\ensuremath{'}}\{(e\mapsto e')*r\}}$
\item Backward reasoning\\
 $\dfrac{}{\{(e\mapsto-)*((e\mapsto e')-\!\!*\,\,p)\}\texttt{ [e]:= e' }\{p\}}$
\end{itemize}
\item Free

\begin{itemize}
\item Global (backward) reasoning \\
$\dfrac{}{\{(e\mapsto-)*r\}\texttt{ free(e) }\{r\}}$
\end{itemize}
\item Allocation

\begin{itemize}
\item Global reasoning (forward) \\
$\dfrac{}{\{r\}\texttt{ x:= cons(\ensuremath{e_{1},\dots,e_{k}}) }\{\exists x'.(x\mapsto\ensuremath{e_{1}[x'/x],\dots,e_{k}[x'/x])*r[x'/x]}\}}$
\item Backward reasoning\\
 $\dfrac{}{\{\forall x'.(x'\mapsto e_{1},\dots,e_{k})-\!\!*\,\,p[x'/x]\}\texttt{ x:= cons(\ensuremath{e_{1},\dots,e_{k}}) }\{p\}}$
\end{itemize}
\item Lookup 

\begin{itemize}
\item Global reasoning\\
$\dfrac{}{\{\exists x''.(e\mapsto x'')*r[x/x']\}\texttt{ x:= [e] }\{\exists x'.(e[x'/x]\mapsto x)*r[x/x'']\}}$\\
here $x$, $x'$ and $x''$ are distinct, $x'$ and $x''$ do not
occur free in $e$, and $x$ is not free in $r$. 
\item Backward reasoning\\
$\dfrac{}{\{\exists x'.(e\mapsto x')*((e\mapsto x')-\!\!*\,\,p[x'/x])\}\texttt{ x:= [e] }\{p\}}$
\end{itemize}
\end{itemize}

\section{\label{sec:Annotated-proofs} Annotated proofs}

\subsubsection*{Proof outlines:}

We have already used assertions in Hoare-triples to state what is
true before and after the execution of an instruction. In a similar
way, an assertion can also be inserted between any two commands of
a program to state what must be true at that point of execution. Placing
assertions in this way is also called \emph{annotating} the program.

For example, consider the following annotated program that swaps the
value of variable x and y using a third variable z. Note the use of
$X$ and $Y$ to represent the initial values of variable $x$ and
$y$ respectively.

\begin{center}
\begin{tabular}{l}
$\{x=X\wedge y=Y\}$\tabularnewline
$z\texttt{:= }x$;\tabularnewline
$\{z=X\wedge x=X\wedge y=Y\}$\tabularnewline
$x\texttt{:= }y$;\tabularnewline
$\{z=X\wedge x=Y\wedge y=Y\}$\tabularnewline
$y\texttt{:= }z$;\tabularnewline
$\{x=Y\wedge y=X\}$\tabularnewline
\end{tabular}
\par\end{center}

Validity of each Hoare-triple in the above program can easily be checked
using axioms for assignment. Hence one concludes that the program
satisfies its specification.
\begin{itemize}
\item A program together with an assertion between each pair of statement
is called a \emph{fully annotated} program. 
\item One can prove that a program satisfies its specification by proving
the validity of every consecutive Hoare-triple which is present in
its annotated version. Hence, a fully annotated program provides a
complete \emph{proof outline} for the program. 
\end{itemize}
Now, we consider another annotated program that involves assertions
from the separation logic. Note that the assertion $(x\mapsto a,o)*(x+o\mapsto b,-o)$
can be used to describe a circular offset-list. Here is a sequence
of commands that creates such a cyclic structure: 

\begin{center}
\begin{tabular}{cl}
1 & $\{\textbf{emp}\}$\tabularnewline
 & $x\texttt{:= cons}(a,a)$;\tabularnewline
2 & $\{x\mapsto a,a\}$\tabularnewline
 & $t\texttt{:= cons}(b,b)$;\tabularnewline
3 & $\{(x\mapsto a,a)*(t\mapsto b,b)\}$\tabularnewline
 & $[x+1]\texttt{:= }t-x$;\tabularnewline
4 & $\{(x\mapsto a,t-x)*(t\mapsto b,b)\}$\tabularnewline
 & $[t+1]\texttt{:= }x-t$;\tabularnewline
5 & $\{(x\mapsto a,t-x)*(t\mapsto b,x-t)\}$\tabularnewline
6 & $\{\exists o.(x\mapsto a,o)*(x+o\mapsto b,-o)\}$\tabularnewline
\end{tabular}
\par\end{center}

The above proof outline illustrates two important points. 
\begin{itemize}
\item First, a label is used against each assertion so that referring becomes
easy in the future discussions. 
\item Secondly, the adjacent assertions - e.g. here the assertions 5 and
6 - mean that the first implies the second. 
\end{itemize}
Also, note the use of $*$ in assertions 3. It insures that $x+1$
is different from $t$ and $t+1$, and hence the assignment $[x+1]\texttt{:= }t-x$
cannot affect the $t\mapsto b,b$ clause. A similar reasoning applies
for the last command as well.

\subsubsection*{Inductive definitions:}

When reasoning about programs which manipulate data structure, we
often need to use inductively defined predicates describing such structures.
For example, in any formal setting, if we wish to reason about the
contents of a linked list we would like to relate it to the abstract
mathematical notion of sequences.
\begin{itemize}
\item Consider the following inductive definition of a predicate that describes
the content of a linked list
\end{itemize}
\begin{center}
\begin{tabular}{rcl}
$\textsf{listrep }\epsilon\,\,(i,j)$ & $\triangleq$ & $i=j\wedge\textbf{ emp}$\tabularnewline
$\textsf{listrep }\text{a}.\alpha\,\,(i,k)$ & $\triangleq$ & $i\neq j\,\,\wedge\,\,\exists j.i\mapsto\text{a},j*\textsf{listrep }\alpha\,\,(j,k)$\tabularnewline
\end{tabular}
\par\end{center}

Here $\alpha$ denotes a mathematical sequence. Informally, the predicate
$\textsf{listrep }\alpha\,\,(x,y)$ claims that $x$ points to a linked
list segment ending at $y$ and the contents (head elements) of that
segment are the sequence $\alpha$. 
\begin{itemize}
\item While proving programs in this section we use $x\overset{\alpha}{\rightsquigarrow}y$
as an abbreviation for $\textsf{listrep \ensuremath{\alpha}\,\,\ensuremath{(x,y)}}$
and $\alpha^{\dagger}$ to represent the reverse of the sequence $\alpha$.
\end{itemize}

\subsubsection*{Proof of In-place list reversal:}

Consider the following piece of code, that performs an in-place reversal
of a linked list: 

\begin{center}
\begin{tabular}{lc}
\{$i\ensuremath{\overset{\alpha_{0}}{\rightsquigarrow}}\boxtimes$\} & \tabularnewline
/{*} i points to the initial linked list {*}/ & \tabularnewline
$\texttt{j:= }\boxtimes$ ; & \tabularnewline
$\texttt{while i}\neq\boxtimes$ do & \tabularnewline
($\texttt{k:= [i+1];}$$\texttt{ [i+1]:= j;}$$\texttt{ j:= i;}$
$\texttt{i:= k ;}$) & \tabularnewline
/{*} j points to the in place reversal of the initial list pointed
by i {*}/ & \tabularnewline
\{$j\ensuremath{\overset{\alpha_{0}^{\dagger}}{\rightsquigarrow}}\boxtimes$\} & \tabularnewline
\end{tabular}
\par\end{center}

Here, $\boxtimes$ represents the null pointer. On a careful analysis
of the code it is easy to see that, 
\begin{itemize}
\item At any iteration of the while loop, variable $i$ and $j$ points
to two different list segments having the contents $\alpha$ and $\beta$
such that concatenating $\beta$ at the end of $\alpha^{\dagger}$
will always result in $\alpha_{0}$. 
\item Thus, we have the following loop invariant 
\end{itemize}
\begin{center}
$\exists\alpha,\beta.\texttt{ (i\ensuremath{\overset{\alpha}{\rightsquigarrow}}\ensuremath{\boxtimes}\ * j\ensuremath{\overset{\beta}{\rightsquigarrow}\boxtimes}) \ensuremath{\wedge}\ \ensuremath{\alpha_{0}^{\dagger}=\alpha^{\dagger}.\beta}}$
\par\end{center}

where, $\alpha_{0}$ represents the initial content of linked list
pointed by variable $i$. 

Also, note the use of separating conjunction in the loop invariant
instead of the usual classical conjunction. If there is any sharing
between the lists i and j then the program may malfunction. The use
of a classical conjunction here cannot not guarantee such non-sharing. 

It is easy to see how the postcondition of the list reversal program
follows from the above loop invariant. Following sequence of specifications
gives a derivation of the postcondition assuming the loop invariant
and the termination condition $i=\boxtimes$,

\begin{center}
\begin{tabular}{cl}
8 & \{$\exists\alpha,\beta.\texttt{ (i\ensuremath{\overset{\alpha}{\rightsquigarrow}}\ensuremath{\boxtimes}\ * j\ensuremath{\overset{\beta}{\rightsquigarrow}\boxtimes}) \ensuremath{\wedge}\ \ensuremath{\alpha_{0}^{\dagger}=\alpha^{\dagger}.\beta}}\wedge\,\,i=\boxtimes$\}\tabularnewline
8a & \{$\exists\beta.\texttt{ (i\ensuremath{\overset{\epsilon}{\rightsquigarrow}}\ensuremath{\boxtimes}\ * j\ensuremath{\overset{\beta}{\rightsquigarrow}\boxtimes}) \ensuremath{\wedge}\ \ensuremath{\alpha_{0}^{\dagger}=\epsilon^{\dagger}.\beta}}$\}\tabularnewline
8b & \{$\exists\beta.\texttt{ (i\ensuremath{\overset{\epsilon}{\rightsquigarrow}}\ensuremath{\boxtimes}\ * j\ensuremath{\overset{\beta}{\rightsquigarrow}\boxtimes}) \ensuremath{\wedge}\ \ensuremath{\alpha_{0}^{\dagger}=\beta}}$\}\tabularnewline
8c & \{$\texttt{ (i\ensuremath{\overset{\epsilon}{\rightsquigarrow}}\ensuremath{\boxtimes}\ * j\ensuremath{\overset{\alpha_{0}^{\dagger}}{\rightsquigarrow}\boxtimes}) }$\}\tabularnewline
8d & \{$\texttt{ (j\ensuremath{\overset{\alpha_{0}^{\dagger}}{\rightsquigarrow}\boxtimes}) }$\}\tabularnewline
\end{tabular}
\par\end{center}

Where, the loop invariant can be verified using the following proof
outline:

\begin{center}
\begin{tabular}{cc||c||c||l||c||l}
1 & \multicolumn{6}{l}{$\{\exists\alpha,\beta.(i\overset{\alpha}{\rightsquigarrow}\boxtimes\,\,*\,\,j\overset{\beta}{\rightsquigarrow}\boxtimes)\wedge\alpha_{0}^{\dagger}=\alpha^{\dagger}.\beta\wedge i\neq\boxtimes\}$}\tabularnewline
2 & \multicolumn{6}{l}{$\{\exists\alpha'.(i\mapsto a,p\,\,*\,\,p\overset{\alpha'}{\rightsquigarrow}\boxtimes\,\,*\,\,j\overset{\beta}{\rightsquigarrow}\boxtimes)\wedge\alpha_{0}^{\dagger}=(a.\alpha')^{\dagger}.\beta\}$}\tabularnewline
 & \multicolumn{6}{l}{$\texttt{k:= [i+1] }$;}\tabularnewline
3 & \multicolumn{6}{l}{$\{\exists\alpha'.(i\mapsto a,k\,\,*\,\,k\overset{\alpha'}{\rightsquigarrow}\boxtimes\,\,*\,\,j\overset{\beta}{\rightsquigarrow}\boxtimes)\wedge\alpha_{0}^{\dagger}=(a.\alpha')^{\dagger}.\beta\}$}\tabularnewline
 & \multicolumn{6}{l}{$\texttt{[i+1]:= j ;}$}\tabularnewline
4 & \multicolumn{6}{l}{$\{\exists\alpha'.(i\mapsto a,j\,\,*\,\,k\overset{\alpha'}{\rightsquigarrow}\boxtimes\,\,*\,\,j\overset{\beta}{\rightsquigarrow}\boxtimes)\wedge\alpha_{0}^{\dagger}=(a.\alpha')^{\dagger}.\beta\}$}\tabularnewline
5 & \multicolumn{6}{l}{$\{\exists\alpha',\beta'.(k\overset{\alpha'}{\rightsquigarrow}\boxtimes\,\,*\,\,i\overset{\beta'}{\rightsquigarrow}\boxtimes)\wedge\alpha_{0}^{\dagger}=(\alpha')^{\dagger}.\beta'\}$}\tabularnewline
 & \multicolumn{6}{l}{$\texttt{j:= i ;}$}\tabularnewline
6 & \multicolumn{6}{l}{$\{\exists\alpha',\beta'.(k\overset{\alpha'}{\rightsquigarrow}\boxtimes\,\,*\,\,j\overset{\beta'}{\rightsquigarrow}\boxtimes)\wedge\alpha_{0}^{\dagger}=(\alpha')^{\dagger}.\beta'\}$}\tabularnewline
 & \multicolumn{6}{l}{$\texttt{i:= k ;}$}\tabularnewline
7 & \multicolumn{6}{l}{$\{\exists\alpha',\beta'.(i\overset{\alpha'}{\rightsquigarrow}\boxtimes\,\,*\,\,j\overset{\beta'}{\rightsquigarrow}\boxtimes)\wedge\alpha_{0}^{\dagger}=(\alpha')^{\dagger}.\beta'\}$}\tabularnewline
7a & \multicolumn{6}{l}{$\{\exists\alpha,\beta.(i\overset{\alpha}{\rightsquigarrow}\boxtimes\,\,*\,\,j\overset{\beta}{\rightsquigarrow}\boxtimes)\wedge\alpha_{0}^{\dagger}=(\alpha)^{\dagger}.\beta\}$}\tabularnewline
\end{tabular}
\par\end{center}

Moreover, the following sequence of assertions gives a detailed proof
of the implications $1\implies2$ and $4\implies5$: 

\begin{center}
\begin{tabular}{cc||c||c||l||c||l}
1 & \multicolumn{6}{l}{$\{\exists\alpha,\beta.(i\overset{\alpha}{\rightsquigarrow}\boxtimes\,\,*\,\,j\overset{\beta}{\rightsquigarrow}\boxtimes)\wedge\alpha_{0}^{\dagger}=\alpha^{\dagger}.\beta\wedge i\neq\boxtimes\}$}\tabularnewline
1a & \multicolumn{6}{l}{$\{\exists a,\alpha',\beta.(i\overset{a.\alpha'}{\rightsquigarrow}\boxtimes\,\,*\,\,j\overset{\beta}{\rightsquigarrow}\boxtimes)\wedge\alpha_{0}^{\dagger}=(a.\alpha')^{\dagger}.\beta\}$}\tabularnewline
1b & \multicolumn{6}{l}{$\{\exists a,\alpha',\beta,p.(i\mapsto a,p\,\,*\,\,p\overset{\alpha'}{\rightsquigarrow}\boxtimes\,\,*\,\,j\overset{\beta}{\rightsquigarrow}\boxtimes)\wedge\alpha_{0}^{\dagger}=(a.\alpha')^{\dagger}.\beta\}$}\tabularnewline
2 & \multicolumn{6}{l}{$\{\exists\alpha'.(i\mapsto a,p\,\,*\,\,p\overset{\alpha'}{\rightsquigarrow}\boxtimes\,\,*\,\,j\overset{\beta}{\rightsquigarrow}\boxtimes)\wedge\alpha_{0}^{\dagger}=(a.\alpha')^{\dagger}.\beta\}$}\tabularnewline
 & \multicolumn{6}{l}{}\tabularnewline
4 & \multicolumn{6}{l}{$\{\exists\alpha'.(i\mapsto a,j\,\,*\,\,k\overset{\alpha'}{\rightsquigarrow}\boxtimes\,\,*\,\,j\overset{\beta}{\rightsquigarrow}\boxtimes)\wedge\alpha_{0}^{\dagger}=(a.\alpha')^{\dagger}.\beta\}$}\tabularnewline
4a & \multicolumn{6}{l}{$\{\exists\alpha'.(k\overset{\alpha'}{\rightsquigarrow}\boxtimes\,\,*\,\,i\mapsto a,j\,\,*\,\,j\overset{\beta}{\rightsquigarrow}\boxtimes)\wedge\alpha_{0}^{\dagger}=(a.\alpha')^{\dagger}.\beta\}$}\tabularnewline
4b & \multicolumn{6}{l}{$\{\exists\alpha'.(k\overset{\alpha'}{\rightsquigarrow}\boxtimes\,\,*\,\,i\overset{a.\beta}{\rightsquigarrow}\boxtimes)\wedge\alpha_{0}^{\dagger}=(\alpha')^{\dagger}.a.\beta\}$}\tabularnewline
5 & \multicolumn{6}{l}{$\{\exists\alpha',\beta'.(k\overset{\alpha'}{\rightsquigarrow}\boxtimes\,\,*\,\,i\overset{\beta'}{\rightsquigarrow}\boxtimes)\wedge\alpha_{0}^{\dagger}=(\alpha')^{\dagger}.\beta'\}$}\tabularnewline
\end{tabular}
\par\end{center}
\begin{itemize}
\item Explanations: 

\begin{itemize}
\item Most of the proof steps, specially those around a command, comprises
of Hoare triples, which can easily be verified using the axioms for
the corresponding commands.
\item Note the use of $*$ instead of $\wedge$ in assertion 3. It insures
that $i+1$ is different from $k$ and $j$. Hence an attempt to mutate
the location $i+1$ does not affect the remaining two clauses $k\overset{\alpha'}{\rightsquigarrow}\boxtimes\,\,$
and $\,\,j\overset{\beta}{\rightsquigarrow}\boxtimes$.
\item We can obtain 1a from 1 by using definition of $i\overset{\alpha}{\rightsquigarrow}\boxtimes$
with the fact that $i\neq\boxtimes$. Then we unfold the definition
of $i\overset{a.\alpha'}{\rightsquigarrow}\boxtimes$ to obtain 1b
from 1a. Finally instantiating $\exists a$ takes us to 2. 
\item 4a is a simple rearrangement of 4. Since $i\overset{a.\beta}{\rightsquigarrow}\boxtimes$
is a shorthand for $i\mapsto a,j\,\,*\,\,j\overset{\beta}{\rightsquigarrow}\boxtimes$
we can obtain 4b from 4a. Finally we obtain 5 by generalizing $a.\beta$
in 4b as $\beta'$ using the existential quantifier.
\end{itemize}
\end{itemize}

\section{Conclusion}

In this article we reviewed some of the important features of Separation
Logic, that first appeared in \cite{key-4,key-5,key-7}. The key idea
of separating conjunction was inspired by the Burstall's \cite{key-2}
``distinct non-repeating tree systems''. It is based on the idea
of organizing assertions to localize the effect of a mutation. The
separating conjunction gives us a succinct and more intuitive way
to describe the memory disjointness, when pointers are involved. However,
it is not the only possible way. One can see \cite{key-8} for references
and other works on proving pointer programs using the standard Hoare
Logic. 

In this paper we considered simple data structures to illustrate the
power of Separation Logic. A more elaborate discussion with a variety
of data structures can be found in \cite{key-7,key-9}. Reasoning
becomes difficult when data structure uses more sharing. In this direction,
one can refer Yang's proof \cite{key-11} of the Schorr- Waite graph
marking algorithm. 

We did not talk much about the proof theory behind Separation Logic.
For a detailed discussion on the soundness and completeness results,
one can refer \cite{key-3,key-6}. The soundness results for most
of the derived rules, presented in this paper, can also be found in
\cite{key-9}.

Finally, it should be noted that the goal here is not to identify
a sound and complete logic for program verification. Instead, the
challenge is to come up with a formalism that can capture the informal
local reasoning used by programmers. Programmers often assume non
sharing between data structures, which need explicit mention when
using standard techniques, such as Hoare Logic. On the other hand,
memory disjointness is default in the separating conjunction. Hence,
separation logic gives us a more natural and concise way to model
a programmer's reasoning.

\end{document}